\documentclass{emulateapj}
\usepackage{bm}

\newcommand{\plotbig}[3] 
{\begin{figure*}\epsscale{#3}
\plotone{#1.eps}
\caption{#2 \label{fig:#1}}
\end{figure*}}

\def\etal{et al.\ }

\def\eg{e.\,g.\,}

\def\unit #1{\,{\rm #1}} 
\def\micron{\unit{\mu m}}
\def\arcsec{\unit{arcsec}}

\def\arcsecsqi{\unit{arcsec^{-2}}}
\def\arcmin{\unit{arcmin}}
\def\mjy{\unit{mJy}}

\def\magarcsecsqi{\unit{mag\arcsecsqi}}

\def\halpha{{\rm H\,\alpha }}
\def\hubble#1{H_0={#1}\unit{km\,sec^{-1}\,Mpc^{-1}}}

\shorttitle{Multicolor Surface Photometry of Lenticulars}
\shortauthors{Barway \etal}
\begin{document}
\title{
Multicolor Surface Photometry of Lenticulars I. The Data}

\author{
Sudhanshu Barway\altaffilmark{1},
Y. D. Mayya\altaffilmark{2},
Ajit K. Kembhavi\altaffilmark{3}, and
S. K. Pandey\altaffilmark{1}
}

\altaffiltext{1}{School of Studies in Physics,
Pt. Ravishankar Shukla University, Raipur 492010,India; ircrsu@sancharnet.in}
\altaffiltext{2}{Instituto Nacional de Astrofisica, Optica y Electronica,
Luis Enrique Erro 1, Tonantzintla, Apdo Postal 51 y 216, C.P. 72000,
Puebla, M\'exico; ydm@inaoep.mx}
\altaffiltext{3}{Inter University center for Astronomy and Astrophysics,
Post Bag 4, Ganeshkhind, Pune 411 007, India;
 akk@iucaa.ernet.in}

\begin{abstract}

We present in this paper multicolor surface and aperture photometry in
the $B,  V, R$ and $K'$ bands  for a sample of  34 lenticular galaxies
from the UGC catalogue.   From surface photometric analysis, we obtain
radial profiles  of surface brightness,  colors, ellipticity, position
angle  and the Fourier  coefficients which  describe the  departure of
isophotal shapes from purely elliptical  form and find the presence of
dust lanes, patches and ring like structure in several galaxies in the
sample.   We obtain total  integrated magnitudes  and colors  and find
that  these  are  in good  agreement  with  the  values from  the  RC3
catalogue. Isophotal colors are  correlated with each other, following
the sequence expected for  early-type galaxies. The color gradients in
lenticulars  are more  negative  than the  corresponding gradients  in
ellipticals. There is a good correlation between $B-V$ and $B-R$ color
gradients, and the  mean gradient in the $B-V,  B-R$ and $V-K'$ colors
are $-0.13\pm0.06$,  $-0.18\pm0.06$, $-0.25\pm0.11$ magnitude  per dex
in radius respectively.

\end{abstract}


\section{Introduction}
   
Hubble (1936)  introduced lenticular (S0) galaxies  as a morphological
transition  class  between ellipticals  and  early-type spirals.   The
lenticular  galaxies have disks  with luminosity  ranging from  ten to
hundred percent  of the bulge luminosity, but  without any conspicuous
spiral arms.   From their appearance, and also  their stellar content,
they seem  to be more like  ellipticals rather than  spirals, and have
often been misclassified due to this fact.

It  has  been suggested  (van  den Bergh  1994)  that  there could  be
different, but  overlapping, sub-populations amongst  the lenticulars.
There  are  several scenarios  possible  for  the  formation of  these
galaxies.   They could  be of  primordial origin,  or could  have been
formed  by  the stripping  of  gas  from  spirals, which  changes  the
morphology (Abadi,  Moore \&  Bower 1999), or  through the  mergers of
unequal-mass  spirals  (Bekki 1998).   There  could  be a  significant
difference in the ages of disks  and bulges, in which case they should
have different stellar population, leading to different colors. Bothun
\& Gregg (1990)  found that bulges and disks  of lenticulars were well
separated in the $B-H$ vs $J-K$ diagram which they interpreted to mean
that the  disks are younger  than bulges by  3-5 Gyr. But  Peletier \&
Balcells (1996) found that the color differences between the disks and
bulges are  much smaller than those  found by Bothun  \& Gregg (1990),
and inferred  that the  disks are only  slightly younger  than bulges,
with the difference in age being in  the range 0-3 Gyr.  In a study of
12 highly  inclined lenticulars, Michard \& Poulain  (2000) found that
disks were often redder than  bulges, which indicates that there could
be large  concentration of  dust in the  disks.  A  detailed multiband
study of the morphology of representative samples of lenticulars, with
possible separation  of bulge and  disk components, and  comparison of
their properties  with those of  ellipticals, and with the  bulges and
disks of spirals will be  very important in addressing these and other
possibilities.

We have observed a sample of 34 lenticular galaxies in the optical and
infrared bands, and  present in this paper their  surface and aperture
photometry, together with color profiles and mean color gradients.  We
will  in  a subsequent  publication  present  a  decomposition of  the
surface brightness  distribution of the  galaxies into bulge  and disk
components using  the decomposition technique  developed by Wadadekar,
Robbason \&  Kembhavi (1999). Using the results  of the decomposition,
it  should be  possible  to  see where  the  lenticulars are  situated
relative to  the fundamental plane of ellipticals  (see \eg Jorgensen,
Franx \&  Kjaergaard 1996 and references therein)  and the photometric
plane for ellipticals and  the bulges of spiral galaxies (Khosroshahi,
Wadadekar \& Kembhavi 2000a and 2000b).  With multiband data, it might
be possible  to determine in this manner  the bulge-to-disk luminosity
ratios  of  these  galaxies  in   different  bands,  as  well  as  the
distribution of  different colors as  a function of distance  from the
center in the bulge  and disk components separately, thus facilitating
the comparison  with other  galaxy types.  Our  data set will  be very
useful for detailed comparison with  other galaxy types, as well as in
constraining stellar population models for the lenticulars.

The  present paper  deals with  the details  of the  sample selection,
observations, data reduction and  analysis. This paper is organized as
follows  :   the  selection  criteria  for  the   sample,  details  of
observations and the technique used in data reduction are described in
$\S$2.   The analysis  and  results from  the  surface photometry  and
aperture photometry are discussed in $\S$3.  The distribution of color
gradients  are discussed  in $\S$4  and a  summary of  the  results is
presented  in  $\S$5.  An  appendix  contains  comments on  individual
galaxies.

\begin{deluxetable*}{ccccccc}
\tabletypesize{\small} 
\tablewidth{0pt} 
\tablecaption{Basic parameters of  the  sample  galaxies}  
\tablehead{  
\multicolumn{2}{c}{Galaxy}  &
\colhead{Type} & \colhead{Diameter}  & \colhead{$z$} & \colhead{$m_{ph}$} &
\colhead{$M_B$}   \\  
\colhead{}  &   \colhead{}  &   \colhead{($T$)}  &
\colhead{(arcmin)}  & \colhead{} & \colhead{} & \colhead{} } 
\startdata
UGC 00080 & NGC  0016  & $-$3.0 & 1.8x1.0  & 0.010194 & 12.50 & $-$20.93 \\
UGC 00491 & NGC  0252  & $-$1.0 & 1.5x1.1  & 0.016645 & 13.40 & $-$21.65 \\
UGC 00859 & NGC  0473  & ~~~0.0 & 1.7x1.1  & 0.007118 & 13.20 & $-$19.82 \\
UGC 00926 & NGC  0499  & $-$2.5 & 1.6x1.3  & 0.014673 & 13.00 & $-$21.55 \\
UGC 01250 & NGC  0670  & $-$2.0 & 2.0x1.0  & 0.012352 & 13.10 & $-$20.76 \\
UGC 01823 & NGC  0890  & $-$3.0 & 2.5x1.7  & 0.013323 & 12.50 & $-$22.31 \\
UGC 01964 & NGC  0940  & $-$2.0 & 1.2x1.0  & 0.017319 & 13.40 & $-$21.64 \\
UGC 02039 & NGC  0969  & $-$2.0 & 1.7x1.6  & 0.015054 & 13.50 & $-$21.51 \\
UGC 02187 & NGC  1040  & $-$2.0 & 1.7x0.8  & 0.016054 & 14.00 & $-$21.06 \\
UGC 02322 & NGC  1106  & $-$1.0 & 1.8x1.8  & 0.014467 & 13.70 & $-$21.39 \\
UGC 03087 &            & $-$2.0 & 0.8x0.6  & 0.033010 & 14.20 & $-$22.28 \\ 
UGC 03178 & NGC  1671  & $-$2.0 & 1.1x0.9  & 0.021270 & 13.90 & $-$21.63 \\ 
UGC 03452 & NGC  2208  & $-$2.0 & 1.7x1.0  & 0.018763 & 14.00 & $-$21.26  \\ 
UGC 03536 &            & $-$2.0 & 1.1x0.6  & 0.015641 & 14.40 & $-$20.86 \\  
UGC 03567 &            & $-$2.0 & 1.0x0.8  & 0.020164 & 14.20 & $-$21.26 \\  
UGC 03642 &            & $-$2.0 & 1.5x1.1  & 0.015004 & 13.50 & $-$21.47 \\ 
UGC 03683 &            & $-$2.0 & 2.0x1.3  & 0.019076 & 14.10 & $-$21.53 \\  
UGC 03699 & NGC  2332  & $-$2.0 & 1.5x1.0  & 0.019467 & 14.00 & $-$21.52 \\ 
UGC 03792 &            & ~~~0.0 & 1.8x1.3  & 0.020608 & 14.00 & $-$21.46 \\ 
UGC 03824 &            & $-$2.0 & 0.9x0.8  & 0.017866 & 14.40 & $-$20.76 \\
UGC 04347 & NGC  2563  & $-$2.0 & 2.1x1.5  & 0.014944 & 13.70 & $-$21.06 \\
UGC 04767 &            & $-$2.0 & 1.3x1.1  & 0.024127 & 14.00 & $-$21.80 \\ 
UGC 04901 & NGC  2804  & $-$2.0 & 2.2x2.0  & 0.028099 & 14.00 & $-$22.23 \\ 
UGC 05292 & NGC  3032  & $-$2.0 & 2.0x1.8  & 0.005114 & 13.00 & $-$19.43 \\ 
UGC 06013 &            & $-$2.0 & 0.9x0.8  & 0.021898 & 13.90 & $-$21.69 \\ 
UGC 06389 & NGC  3648  & $-$2.0 & 1.3x0.8  & 0.006631 & 13.50 & $-$19.50 \\ 
UGC 06899 & NGC  3971  & $-$2.0 & 1.4x1.2  & 0.022489 & 13.90 & $-$21.75 \\
UGC 07142 & NGC  4143  & $-$2.0 & 2.3x1.4  & 0.003616 & 12.00 & $-$20.03 \\ 
UGC 07473 & NGC  4350  & $-$2.0 & 3.0x1.4  & 0.004140 & 11.50 & $-$20.48 \\ 
UGC 07880 & NGC  4638  & $-$3.0 & 2.2x1.4  & 0.003883 & 12.20 & $-$19.64 \\
UGC 07933 & NGC  4673  & $-$5.0 & 1.0x0.9  & 0.022856 & 13.70 & $-$21.99 \\ 
UGC 08675 & NGC  5273  & $-$2.0 & 2.8x2.5  & 0.003616 & 12.70 & $-$19.24 \\ 
UGC 09200 & NGC  5580  & $-$2.0 & 1.8x1.8  & 0.010814 & 13.60 & $-$20.46 \\ 
UGC 09592 & NGC  5784  & $-$2.0 & 1.9x1.8  & 0.017912 & 13.70 & $-$21.78 \\ 
UGC 11178 & NGC  6599  & $-$2.0 & 1.3x1.2  & 0.010120 & 13.70 & $-$20.30 \\ 
UGC 11356 & NGC  6703  & $-$2.5 & 2.5x2.3  & 0.008209 & 12.40 & $-$21.14 \\  
UGC 11781 &            & $-$2.0 & 1.4x1.1  & 0.015501 & 13.70 & $-$21.74 \\ 
UGC 11972 & NGC  7248  & $-$2.5 & 1.7x0.9  & 0.014627 & 13.60 & $-$21.22 \\ 
UGC 12443 & NGC  7539  & $-$2.0 & 1.5x1.2  & 0.020174 & 13.70 & $-$21.88 \\  
UGC 12655 &            & $-$2.0 & 1.4x0.9  & 0.017269 & 14.00 & ~$-$21.24 
\enddata 
\tablecomments{Columns  (1) and
(2) give   the  UGC   catalogue  number   and  NGC   catalogue  number
respectively, column (3) gives the  morphological type from the RC3
catalogue,  columns (4) and  (5)  give  the diameter (in arcmin) along the 
major and minor axis respectively, and 
the  redshift from NASA Extragalactic Database (NED), column (6) give the
photographic magnitude from the UGC catalogue  and column (7) give
the absolute B magnitude for $\hubble{50}$.}
\end{deluxetable*} 

\section{Observations and data reduction}

\subsection{The sample}

Our sample consists of 40 bright and medium sized galaxies, classified
as lenticulars, from the  Uppsala General Catalogue of Galaxies (UGC).
The UGC is  essentially complete to a limiting  major-axis diameter of
one $\arcmin$ and/or  to a limiting apparent magnitude  of 14.5 on the
blue prints of the Palomar Observatory Sky Survey (POSS).  Coverage is
limited to the sky north  of declination $-2.5^\circ$.  For our sample
we selected galaxies with apparent blue magnitude brighter than $m_B =
14$, diameter $D_{\rm 25} < 3\arcmin$ and declination in the range $5<
\delta <  64^\circ$.  The sample galaxies  are listed in  Table 1. Our
sample, though not complete,  is representative of lenticular galaxies
in the field with the above properties.

\plotbig{sb.fig1}{Gray-scale representation of color composite images of 
our sample galaxies (see $\S$ 2.4).}{1}

\subsection{Optical imaging}

Our  optical $BVR$-band observations  were carried  out with  the {\it
Observatorio  Astrof{\'{\i}}sico Guillermo  Haro}  2.1-m telescope  at
Cananea, Mexico.  We used a  Tektronics CCD of $1024\times 1024$ pixel
format at the  $f/12$ Cassegrain focus of the  telescope with $3\times
3$ pixel binning, resulting in  a scale of $0\farcs6$ pixel$^{-1}$ and
a field  of view of $3\farcm4\times3\farcm4$.  Typically two exposures
each  of  ten,  five  and  five  minutes in  $B,  V$,  and  $R$  bands
respectively were  taken for each galaxy. Exposure  times were reduced
if  the  central  pixels  were  close  to  saturation.   Twilight  sky
exposures were taken for  flat-fielding purposes.  Several bias frames
were  obtained  at  the  start   and  end  of  each  night.   All  the
observations were carried out in three runs in December 2001, February
2002 and  October 2002. A  total of 34  galaxies were observed  in the
optical bands. For  the galaxies UGC\,1964, 3178, 4901,  6013 and 7933
from the observed  set, the quality of the images  was not good enough
for  photometric analysis.   Table 2  contains a  detailed log  of the
observations.

\subsection{Near infrared imaging}

We obtained images of all 40 sample galaxies in the near infrared $K'$
band with the {\it  Observatorio Astronomico Nacional} 2.1-m telescope
at  San Pedro  Martir, Mexico.   The CAMILA  instrument (Cruz-Gonzalez
\etal 1994), which  hosts a NICMOS 3 detector  of 256$\times$256 pixel
format,  was  used  in  the   imaging  mode  with  the  focal  reducer
configuration  $f/4.5$ in  all our  observations.  This  results  in a
spatial resolution  of $0\farcs85$ pixel$^{-1}$  and a total  field of
view of $3\farcm6\times3\farcm6$.   Each $K'$ observation consisted of
a sequence of  object and sky exposures, with  the integration time of
an individual exposure limited by the sky counts (or in some cases the
nucleus),  which was  kept well  below  the non-linear  regime of  the
detector.  A  typical $K'$ image  sequence consisted of  10 exposures,
six on  the object and  four on the  sky. The net exposure  times were
typically of  10 minutes.  A  series of twilight and  night-sky images
were  taken for  flat-fielding  purposes. The  $K'$ observations  were
carried out  in four runs in  December 2000, March  2001, October 2001
and March 2002.
 
The  sky conditions for  both optical  and near  infrared observations
were generally photometric and the seeing \emph{FWHM} was in the range
1$\farcs5$--2$\farcs5$ on different nights. The average sky brightness
was 21.14, 20.72,  20.18 and 12.27 magnitude arcsec$^{-2}$  in the $B,
V, R$  and $K'$  bands, respectively. The  sky brightness in  the $K'$
band also includes the background emitted by the warm optics.

\begin{deluxetable*}{ccccccccccc}
\tabletypesize{\small}
\tablewidth{0pt}    
\tablecaption{Log    of observations} 
\tablehead{  \colhead{Galaxy } & \multicolumn{2}{c}{Date
of   Observation}   &   \multicolumn{4}{c}{Exposure  Time   (sec)}   &
\multicolumn{4}{c}{Seeing   {\it   FWHM}   (arcsec)}\\  
\colhead{}   &
\colhead{$BVR$}   &  \colhead{$K'$}  &   \colhead{$B$}  &   \colhead{$V$}  &
\colhead{$R$} & \colhead{$K'$} & \colhead{$B$} & \colhead{$V$} & \colhead{$R$}
& \colhead{$K'$}  } 
\startdata 
UGC 00080 & 04-10-2002 & 10-10-2001 & 1200 & 360 & 540 & 750 &  2.24 & 2.07 & 2.03 & 1.56 \\ 
UGC 00491 & 05-10-2002 & 12-10-2001 & 1200 & 600 & 600 & 750 &  1.85 & 2.18 & 2.18 & 1.52 \\ 
UGC 00859 & 14-12-2001 & 09-10-2001 & 1200 & 600 & 600 & 900 &  2.41 & 2.81 & 2.64 & 1.76 \\ 
UGC 00926 & 05-10-2002 & 10-10-2001 & 1200 & 360 & 540 & 750 &  1.80 & 1.66 & 1.78 & 1.49 \\  
UGC 01250 & 04-10-2002 & 10-10-2001 & 1200 & 600 & 600 & 900 &  2.31 & 2.10 & 2.01 & 1.48 \\ 
UGC 01823 & 14-12-2001 & 09-10-2001 & 1200 & 600 & 600 & 750 &  2.43 & 2.78 & 2.67 & 1.61 \\ 
UGC 01964 & 14-12-2001 & 09-10-2001 & 1200 & 600 & 600 & 900 &  2.10 & 2.09 & 2.18 & 1.60 \\  
UGC 02039 & 05-10-2002 & 12-10-2001 & 1200 & 600 & 600 & 600 &  1.91 & 2.11 & 2.01 & 1.64 \\ 
UGC 02187 & ...        & 10-10-2001 & .... & ... & ... & 900 &  ...  & ...  & ...  & 1.47 \\
UGC 02322 & ...        & 09-10-2001 & .... & ... & ... & 780 &  ...  & ...  & ...  & 1.49 \\ 
UGC 03087 & 08-02-2002 & 24-03-2002 & 1200 & 540 & 540 & 480 &  2.16 & 1.98 & 2.14 & 2.06 \\ 
UGC 03178 & 09-02-2002 & 23-03-2002 & 1200 & 600 & 600 & 320 &  2.02 & 1.87 & 1.87 & 1.75 \\  
UGC 03452 & 14-12-2001 & 16-12-2000 & 1200 & 600 & 600 & 600 &  2.55 & 2.76 & 2.63 & 1.58 \\ 
UGC 03536 & 08-02-2002 & 26-03-2002 & 1200 & 540 & 540 & 105 &  2.14 & 2.01 & 2.05 & 1.54 \\ 
UGC 03567 & 07-02-2002 & 28-02-2002 & 1200 & 600 & 600 & 750 &  2.17 & 1.97 & 2.04 & 1.92 \\  
UGC 03642 & 09-02-2002 & 26-02-2002 & 1200 & 720 & 720 & 750 &  1.86 & 2.04 & 1.90 & 1.48 \\ 
UGC 03683 & 08-02-2002 & 23-03-2002 & 1200 & 600 & 600 & 750 &  2.26 & 2.26 & 1.97 & 2.07 \\ 
UGC 03699 & 05-10-2002 & 12-10-2001 &  600 & 360 & 540 & 750 &  2.58 & 2.58 & 2.59 & 1.89 \\  
UGC 03792 & 14-12-2001 & 10-10-2001 & 1200 & 600 & 600 & 750 &  2.28 & 2.69 & 2.43 & 1.85 \\ 
UGC 03824 & 07-02-2002 & 26-03-2002 & 1800 & 600 & 600 & 105 &  2.13 & 2.13 & 2.24 & 1.66 \\ 
UGC 04347 & 14-12-2001 & 08-03-2001 & 1200 & 600 & 600 & 540 &  2.51 & 2.71 & 2.55 & 1.88 \\  
UGC 04767 & 08-02-2002 & 24-03-2002 & 1200 & 600 & 600 & 600 &  2.42 & 2.07 & 2.08 & 1.92 \\ 
UGC 04901 & 09-02-2002 & 26-03-2002 & .... & 600 & 600 & 105 &  ...  & 1.75 & 1.85 & 1.45 \\ 
UGC 05292 & ...        & 07-06-2001 & .... & ... & ... & 900 &  ...  & ...    & ...  & 1.67 \\ 
UGC 06013 & 07-02-2002 & 08-02-2001 & .... & 600 & 600 & 840 &  ...  & 2.50 & 2.07 & 1.64  \\ 
UGC 06389 & 08-02-2002 & 23-03-2002 & 1200 & 540 & 540 & 700 &  1.92 & 2.05 & 2.03 & 1.89 \\ 
UGC 06899 & 08-02-2002 & 23-03-2002 & 1200 & 600 & 600 & 840 &  1.96 & 1.97 & 1.94 & 1.27 \\ 
UGC 07142 & ...        & 23-03-2002 & .... & ... & ... & 450 &  ...  & ...  & ...    & 2.07  \\ 
UGC 07473 & 09-02-2002 & 22-03-2002 &  900 & 450 & 270 & 260 &  2.12 & 2.13 & 2.08 & 1.43 \\ 
UGC 07880 & 09-02-2002 & 25-03-2002 & 1200 & 270 & 600 & 500 &  2.37 & 2.64 & 2.41 & 1.79 \\  
UGC 07933 & 07-02-2002 & 08-03-2002 & .... & 540 & 540 & 840 &  ...  & 2.47 & 2.26 & 1.79 \\ 
UGC 08675 & 08-02-2002 & 22-03-2002 & 1200 & 600 & 540 & 310 &  2.04 & 1.91 & 1.82 & 1.63 \\ 
UGC 09200 & ...        & 07-03-2001 & .... & ... & ... & 840 &  ...  & ...  & ...  & 1.50 \\ 
UGC 09592 & ...        & 24-03-2002 & .... & ... & ....& 600 &  ...  & ...  & ...  & 1.92 \\ 
UGC 11178 & 04-10-2002 & 22-03-2002 &  900 & 600 & 600 & 240 &  2.15 & 2.06 & 2.20 & 1.39 \\  
UGC 11356 & 05-10-2002 & 23-03-2002 & 1200 & 240 & 360 & 420 &  1.91 & 1.91 & 1.91 & 1.80 \\ 
UGC 11781 & 04-10-2002 & 09-10-2001 &  900 & 600 & 600 & 540 &  1.99 & 1.94 & 2.07 & 1.83 \\ 
UGC 11972 & 05-10-2002 & 10-10-2001 &  900 & 540 & 360 & 900 &  2.07 & 1.79 & 1.78 & 1.50  \\  
UGC 12443 & 14-12-2001 & 12-10-2001 & 1200 & 600 & 600 & 900 &  2.87 & 2.89 & 2.92 & 1.56 \\ 
UGC 12655 & 04-10-2002 & 09-10-2001 & 1200 & 600 & 600 & 120 &  2.24 & 1.98 & 2.12 & 1.59 
\enddata
\end{deluxetable*}

\subsection{Data reduction}

The basic data  reduction for both the optical  and near infrared $K'$
images involved  subtraction of the  bias and sky frames,  division by
flat field frames, registration of  the images to a common co-ordinate
system and  then stacking  all the  images of a  given galaxy  in each
filter.  Night  to night  variations of the  optical bias  frames were
negligible,  and hence  bias  frames  of an  entire  run were  stacked
together using the median algorithm to form a master bias frame, which
was then  subtracted from  all the other  frames.  Preparation  of the
optical flat fields followed  the conventional technique, wherein bias
subtracted flats  were stacked and the resultant  frame was normalized
to  its  mean value  to  form  a master  flat  in  each filter.   Bias
subtracted  images  of  the  program  galaxies  were  divided  by  the
normalized flat field in the corresponding filter.  The optical images
suffered from a stray light problem that resulted in a gradient in the
sky background, which roughly ran  through one of the diagonals of the
CCD chip. The gradient was found  to be stable throughout each run and
the  mean counts scaled  linearly with  exposure time.   After several
experiments, we found that the best way to get rid of the gradient was
to subtract  a mean blank  sky image from  the data images.   For this
purpose,  special  blank fields  were  observed  in  each filter  with
exposure  times matching  the  typical exposure  times  of the  object
frames. In the December 2001 run,  three blank sky frames were used in
each filter to prepare a  median gradient image, while in the February
2002 run, six frames were  used.  The adopted procedure eliminated any
gradient from the sky background.

For the $K'$  images, a bias frame taken  immediately before an object
exposure was subtracted as part of the data acquisition. A master $K'$
flat field  for each night of  observing was prepared  as follows. The
night-sky flats  were first stacked  and then subtracted  from stacked
twilight flats.  The frames obtained in this fashion for each run were
then combined and normalized to  the mean value of the resultant frame
to  form a  master $K'$  flat.   The sky  frames of  each sequence  of
observations were combined and the resultant image was subtracted from
each of the object frames to get a sky-subtracted image. Flat fielding
was done  by dividing the sky  subtracted images of the  object by the
normalized master flat.  The resulting images were aligned to a common
co-ordinate system using common stars  in the frames and then combined
using  the median  operation.  Only  good  images (as  defined in  the
CAMILA manual  --- see  Cruz-Gonzalez et al.   1994) were used  in the
$K'$ combination.  The resulting  combined $K'$ images were aligned to
corresponding images from the Digitized  Sky Survey (DSS).  As a final
step of  the reduction procedure, the mutually  aligned optical images
were  aligned to the  $K'$ image  coordinate system.   The transformed
star positions  in the  images agreed to  within $0\farcs2$  as judged
from the coordinates of common stars.

All image  reductions were carried  out using the Image  Reduction and
Analysis  Facility  (IRAF\footnote {IRAF  is  distributed by  National
Optical Astronomy Observatories, which are operated by the Association
of  Universities for  Research in  Astronomy, Inc.,  under cooperative
agreement  with  the National  Science  Foundation.})   and the  Space
Telescope  Science Data  Analysis System  (STSDAS). The  IRAF external
package  {\it color}  was  used  to make  color  composite images,  by
combining images taken in three different bands, $B, V$ and $R$ or $V,
R$ and $K'$. Gray-scale representation of these color composite images
are shown in Fig. 1, where North is up and East is to the left.

\subsection{Photometric calibration}

Dipper  Asterism stars  in the  M\,67  field were  observed to  enable
accurate  photometric calibration  of our  optical  observations.  The
stars in this field span a wide  color range ($-0.05 < B - V < 1.35$),
which includes the range of  colors of the program galaxies, and hence
are  suitable for  obtaining  the transformation  coefficients to  the
Cousins $BVR$  system defined by Bessell  (1990).  The transformation
equations are

%
\begin{eqnarray}
B & = b_0 + \alpha_B + \beta_B (b_0-v_0),  \\
V & = v_0 + \alpha_V + \beta_V (b_0-v_0), \\
R & = r_0 + \alpha_R + \beta_R (v_0-r_0),
\end{eqnarray}
%
where $B, V$ and $R$ are standard magnitudes, $b_0, v_0$ and $r_0$ are
extinction corrected instrumental magnitudes, $\alpha_B, \alpha_V$ and
$\alpha_R$ are  the zero points  and $\beta_B, \beta_V$  and $\beta_R$
the color coefficients  in bands $B, V$ and  $R$ respectively. Typical
extinction coefficients  for the observatory (0.20, 0.11  and 0.07 for
$B, V$  and $R$  bands respectively) were  used. Considering  that the
objects and the standard stars  were observed as close to the meridian
as possible,  and in none of  the cases the airmass  exceeded 1.3, the
error  introduced   due  to  possible  variation   in  the  extinction
coefficients is  less than 0.02 magnitude.   Coefficients $\alpha$ and
$\beta$  were  obtained by  using  the  $BVR$  standard magnitudes  of
Chevalier  \&  Ilovaisky (1991).   Resulting  values  of $\alpha$  and
$\beta$ are given  in Table 3. The relatively  large color coefficient
on the $R$-band  calibration is due to the  non-standard nature of the
filter used in  the observations.  The stability of  the $\alpha$ on a
given night was checked using  at least two other standard fields from
the  Landolt  Selected  Areas  (Landolt 1992).   The  standard  fields
observed during  our runs are  SA\,110$-$232, PG\,2336+004, Rubin\,149
and PG\,1323$-$086.   Overall, the zero  points within a  single night
agreed  to within  0.02 magnitudes.   Night-to-night variation  of the
zero point was also within 0.02 magnitudes.

The detector and  filter system combination that we  used for the near
infrared observations is identical to that used in the observations of
standards by Hunt et al.  (1998), and hence the color coefficients are
expected to be negligibly small.  We verified this by observing fields
AS17 and AS36, which contain stars spanning a wide range of colors. We
observed at least 2 standard  fields each night, each field containing
more  than  one  star  and  some  fields such  as  AS17  containing  5
stars. The  $K'$-band zeropoints  was obtained for  each night  as the
mean  of zero points  from different  stars and  the average  value is
$\alpha_{K'}$ = $20.15\pm0.05$. 

For  reddening corrections  due to  galactic extinction  we  adopt the
values given by  Schlegal \etal (1998) for each  filter in the optical
range. For near infrared observations  we use the extinction law given
by Rieke \etal  (1985).  When galaxies at different  redshift are seen
through  a fixed passband,  the collected  light comes  from different
wavelength   ranges   in   the    rest   frame   of   the   respective
galaxies.  $K$-correction   is  used  to  transform   the  rest  frame
magnitudes of each object to the passband of the filter.  We have used
an approximate  form of the $K$-correction, applicable  for small $z$,
from  Persson  \etal (1979),  Frei  \etal  (1994)  and Fukugita  \etal
(1995). For the four filters we have, $K_{B} = 4.4z$, $K_{V} = 3.24z$,
$K_{R}  = 2.12z$ and  $K_{K'} =  -3z$ respectively,  where $z$  is the
redshift of  the galaxy.  We have listed  the combination  of Galactic
extinction and $K$-correction for individual galaxies in Table 4.

\begin{deluxetable*}{ccc}
\tabletypesize{\small}
\tablewidth{0pt}
\tablecaption{Zero points and transformation coefficients  for optical observations}
\tablehead{
\colhead{Filter}          & \colhead{Zero point} & \colhead{Color coefficient} \\  
\colhead{}              & \colhead{$\alpha$}   & \colhead{$\beta$}                  
}
\startdata
February 2002 &   &   \\ 
\hline \\ 
$B$ &  22.88$\pm$0.03 & $-$0.095$\pm$0.016 \\
$V$ &  23.49$\pm$0.02 & ~~~0.075$\pm$0.010  \\
$R$ &  22.45$\pm$0.05 & $-$0.408$\pm$0.081 \\\\
\hline \\
October 2002 &    &  \\ \hline \\
$B$ &  22.60$\pm$0.03  & $-$0.095$\pm$0.016 \\
$V$ &  23.21$\pm$0.04  & ~~~0.075$\pm$0.010  \\
$R$ &  22.15$\pm$0.03  & $-$0.408$\pm$0.081  
\enddata
\end{deluxetable*}

\section{Analysis}

\subsection{Sky background}

An  accurate estimation of  the sky  background is  a crucial  step in
surface   photometric  analysis,  as   even  small   uncertainties  in
determining the sky  can lead to significant errors  in the estimation
of the galaxy surface  brightness and color profiles, especially where
the  surface brightness  becomes comparable  to or  less than  the sky
brightness.  In  those cases  where the galaxy  image is  small enough
that the frame contains portions  of the sky unaffected by the galaxy,
it is  possible to  estimate the sky  using the ``boxes  method'' (see
Peletier \etal 1990a), where the background is estimated from a series
of boxes chosen  avoiding the sample galaxy, foreground  stars and any
other contaminating  objects which  may be present.   We have  used 20
boxes of size $5\times5$ pixels near the corners of the CCD frame, and
have adopted  the mean of the median  count in these boxes  as the sky
background.  The rms dispersion over the  mean of 20 boxes is found to
be 0.6\%  of the mean in  the $B, V$, and  $R$ bands and  0.2\% of the
mean in the $K'$ band.  When a galaxy is so large that the whole frame
is affected by it, use of  the boxes method leads to overestimation of
the background.   In such cases,  the background can be  determined by
fitting a power-law of the form I(r) = sky + I(0) r$^{-\alpha}$ to the
outer parts  of the  surface brightness profile.   Following Jorgensen
\etal (1992) and  Goudfrooij \etal (1994), we fit  the function to the
region  of the  profile  with  $r >  50''$  typically, separately  for
$\alpha=2$ and 3, and use the mean of the respective best-fit constant
values as the background.  The  error in the sky background estimation
is taken to be half the  difference of the sky values obtained for the
two fits.   We have used both  the techniques for all  galaxies in our
sample, and find that in most cases the two methods agree within $\sim
0.005$  magnitude (i.e.  1$\sigma$ difference),  because of  the small
size of  the galaxies relative  to the CCD  frame. In the case  of the
galaxies UGC\,4347, 7473, 7880,  8675 and UGC\,11356, which cover most
of  the  frame, the  power-law  fit  provides  a background  which  is
$\sim0.06$  magnitude (i.e.  10$\sigma$ difference)  fainter  than the
boxes method,  and we adopt  the former as the  background estimation.
In the $K'$ band images, the background is already subtracted from the
many individual frames for each galaxy during the pre-processing stage
(see $\S 2.4$).   However, to check for any  residual background which
may be present in processed frames used in the analysis, we have again
applied  the two techniques  for the  background estimation.   We find
that the two estimates agree very well ($\sim0.0006$ magnitude, except
in two cases where the agreement is within $\sim0.001$ magnitude).

\begin{deluxetable*}{ccccccccc}
\tabletypesize{\small}
\tablewidth{0pt}
\tablecaption{Integrated magnitudes and colors}
\tablehead{
\colhead{Galaxy }  & \colhead{$B$}  & \colhead{$B-V$} & \colhead{$B-R$} 
& \colhead{$V-K'$} & \multicolumn{4}{c}{Galactic extinction  $ +~K $ correction}\\ 
\colhead{}  & \colhead{}  & \colhead{} & \colhead{} & \colhead{} &\colhead{$B$}  
& \colhead{$B-V$} & \colhead{$B-R$} & \colhead{$V-K'$}   }
\startdata
UGC 00080 & 13.01$\pm$0.04 & 0.95$\pm$0.01 & 1.54$\pm$0.01 & 3.34$\pm$0.04 & 0.25 & 0.06 & 0.10 & 0.20 \\ 
UGC 00491 & 13.46$\pm$0.05 & 0.97$\pm$0.02 & 1.61$\pm$0.01 & 3.36$\pm$0.05 & 0.32 & 0.07 & 0.13 & 0.27 \\ 
UGC 00859 & 13.41$\pm$0.03 & 0.78$\pm$0.01 & 1.28$\pm$0.01 & 3.22$\pm$0.06 & 0.41 & 0.09 & 0.16 & 0.30 \\ 
UGC 00926 & 13.32$\pm$0.05 & 1.02$\pm$0.01 & 1.66$\pm$0.01 & 3.59$\pm$0.03 & 0.36 & 0.08 & 0.14 & 0.29 \\  
UGC 01250 & 13.53$\pm$0.03 & 0.81$\pm$0.01 & 1.37$\pm$0.01 & 3.32$\pm$0.04 & 0.36 & 0.08 & 0.14 & 0.29 \\ 
UGC 01823 & 12.58$\pm$0.03 & 0.96$\pm$0.01 & 1.53$\pm$0.01 & 3.37$\pm$0.04 & 0.39 & 0.09 & 0.15 & 0.31 \\  
UGC 02039 & 13.38$\pm$0.08 & 0.97$\pm$0.02 & 1.59$\pm$0.02 & 3.48$\pm$0.05 & 0.57 & 0.13 & 0.22 & 0.44 \\ 
UGC 03087 & 14.60$\pm$0.07 & 0.62$\pm$0.05 & 1.09$\pm$0.05 & 3.72$\pm$0.18 & 0.42 & 0.10 & 0.17 & 0.35 \\ 
UGC 03452 & 13.70$\pm$0.07 & 1.09$\pm$0.01 & 1.69$\pm$0.01 & 3.69$\pm$0.04 & 0.74 & 0.17 & 0.29 & 0.56 \\ 
UGC 03536 & 13.39$\pm$0.09 & 1.02$\pm$0.01 & 1.60$\pm$0.01 & 3.62$\pm$0.03 & 0.54 & 0.12 & 0.21 & 0.42 \\ 
UGC 03567 & 14.36$\pm$0.06 & 1.03$\pm$0.01 & 1.61$\pm$0.01 & 3.54$\pm$0.05 & 0.44 & 0.10 & 0.18 & 0.36 \\ 
UGC 03642 & 13.51$\pm$0.05 & 0.91$\pm$0.02 & 1.50$\pm$0.02 & 3.39$\pm$0.07 & 0.26 & 0.06 & 0.11 & 0.23 \\ 
UGC 03683 & 13.72$\pm$0.06 & 1.04$\pm$0.01 & 1.64$\pm$0.01 & 3.76$\pm$0.04 & 0.48 & 0.11 & 0.19 & 0.39 \\ 
UGC 03699 & 14.13$\pm$0.12 & 1.08$\pm$0.01 & 1.71$\pm$0.01 & 3.64$\pm$0.02 & 0.45 & 0.10 & 0.18 & 0.37 \\ 
UGC 03792 & 14.07$\pm$0.05 & 1.10$\pm$0.01 & 1.75$\pm$0.01 & 3.57$\pm$0.05 & 0.38 & 0.09 & 0.15 & 0.32 \\ 
UGC 03824 & 14.22$\pm$0.05 & 0.99$\pm$0.01 & 1.57$\pm$0.01 & 3.30$\pm$0.05 & 0.31 & 0.07 & 0.13 & 0.27 \\ 
UGC 04347 & 13.40$\pm$0.04 & 1.01$\pm$0.01 & 1.57$\pm$0.01 & 3.57$\pm$0.04 & 0.25 & 0.06 & 0.10 & 0.22 \\   
UGC 04767 & 14.26$\pm$0.05 & 0.99$\pm$0.01 & 1.53$\pm$0.01 & 3.55$\pm$0.05 & 0.19 & 0.04 & 0.08 & 0.21 \\ 
UGC 06389 & 13.57$\pm$0.03 & 0.93$\pm$0.01 & 1.45$\pm$0.01 & 3.35$\pm$0.05 & 0.11 & 0.02 & 0.04 & 0.10 \\ 
UGC 06899 & 14.01$\pm$0.04 & 0.88$\pm$0.01 & 1.42$\pm$0.01 & 3.37$\pm$0.06 & 0.17 & 0.04 & 0.08 & 0.19 \\ 
UGC 07473 & 12.01$\pm$0.02 & 0.93$\pm$0.01 & 1.46$\pm$0.01 & 3.39$\pm$0.03 & 0.13 & 0.03 & 0.05 & 0.10 \\ 
UGC 07880 & 12.14$\pm$0.03 & 0.91$\pm$0.01 & 1.39$\pm$0.01 & 3.24$\pm$0.04 & 0.12 & 0.03 & 0.05 & 0.09 \\  
UGC 08675 & 12.50$\pm$0.03 & 0.80$\pm$0.02 & 1.31$\pm$0.02 & 3.20$\pm$0.09 & 0.05 & 0.01 & 0.02 & 0.05 \\ 
UGC 11178 & 13.71$\pm$0.06 & 0.89$\pm$0.01 & 1.27$\pm$0.01 & 3.68$\pm$0.04 & 0.70 & 0.16 & 0.27 & 0.51 \\  
UGC 11356 & 12.29$\pm$0.07 & 0.95$\pm$0.01 & 1.53$\pm$0.01 & 3.52$\pm$0.04 & 0.41 & 0.09 & 0.16 & 0.31 \\ 
UGC 11781 & 14.23$\pm$0.07 & 1.18$\pm$0.01 & 1.87$\pm$0.01 & 3.86$\pm$0.03 & 1.12 & 0.26 & 0.43 & 0.82 \\ 
UGC 11972 & 13.90$\pm$0.06 & 1.14$\pm$0.01 & 1.82$\pm$0.01 & 3.79$\pm$0.02 & 0.76 & 0.17 & 0.29 & 0.56 \\   
UGC 12443 & 13.94$\pm$0.04 & 0.99$\pm$0.01 & 1.52$\pm$0.01 & 3.36$\pm$0.05 & 0.52 & 0.12 & 0.21 & 0.42 \\ 
UGC 12655 & 14.09$\pm$0.05 & 0.93$\pm$0.01 & 1.53$\pm$0.01 & 3.50$\pm$0.04 & 0.32 & 0.07 & 0.13 & ~0.27
\enddata 
\tablecomments{Column (2) gives the measured
uncorrected total $B$ magnitude, column (3) to (5) give the measured $B-V$,
 $B-R$, and $V-K'$ color indices, columns (6) to (9) give the combination of Galactic foreground
extinction and $K$ correction.}
\end{deluxetable*}

\subsection{Surface Photometry : Isophotal analysis}

We have fitted ellipses to the  isophotes in our $B,V,R$ and $K'$ band
images,  using the  task ELLIPSE  in the  STSDAS package  available in
IRAF.   The   fitting  algorithm  used  is  described   in  detail  by
Jedrzejewski (1987), and uses the intensity distribution along a trial
ellipse, which can be expressed as a Fourier series
\begin{equation}
I(\phi) = I_0 + \sum_{n}a_n\sin(n\phi) + \sum_{n}b_n\cos(n\phi), 
\end{equation}
where  $\phi$ is  the ellipse  eccentric  anomaly, $I_0$  is the  mean
intensity along the ellipse and $a_n$, $b_n$ are harmonic amplitudes.

The fitting was started a few arcseconds from the center of the galaxy
image to minimize  the effect of seeing, and  stopped at the isophotes
where the  mean count becomes comparable  to three times  the error in
the sky  background.  Stars, bad  pixels etc.  were identified  in the
first round of  the fitting procedure and masked in  the next run. All
the parameters  including the  center of the  ellipse were  allowed to
vary during the fitting.  Variation  in ellipse center was found to be
small ($\sim1.2\arcsec$),  which provides a  check on the  accuracy of
the sky  subtraction. We repeated  the fitting process  with different
modes of  sampling and different starting major-axis  lengths to check
the stability of the extracted parameters.

The   fitting   procedure  provides   the   mean  intensity   (surface
brightness), the ellipticity and  the position angle of the major-axis
of the best fit ellipse, as  a function of the semi-major axis length.
The $B$ band surface brightness profile, and the $B-V, B-R$ and $V-K'$
color profiles are shown for all the galaxies in Fig. 2.  The profiles
of the ellipticity, position angle and the Fourier coefficients $a_3,
a_4,  b_3, b_4$  respectively in  the $B,  R$ and  $K'$ bands  for one
galaxy (UGC\,7880) are  shown in Fig. 3.  Tables  and plots of surface
brightness  and  color  profiles  as  well  as  the  profiles  of  the
ellipticity, position  angle and  the Fourier coefficients  $a_3, a_4,
b_3,  b_4$ respectively  in the  $B,  R$ and  $K'$ bands  for all  the
galaxies     in     the    sample     are     available    at     {\it
http://iucaa.ernet.in/$\sim$sudhan/s0.html}.  Some  description of the
profiles for individual galaxies is provided in the Appendix.
   
Color measurements of galaxies  involving frames with different seeing
full width at  half maximum (\emph{FWHM}) can lead  to errors at small
radii,  and in  particular  within the  seeing  disk which  has to  be
approached  for  measuring nuclear  colors.   These  errors have  been
discussed by Franx \etal (1989), Peletier \etal (1990a) and Goudfrooij
\etal (1994), who have derived  a cutoff radius for color profiles and
discarded colors measured inside this limit, which is usually $2\times
FWHM$.  Idiart \etal  (2002)  used  a method  of  equalization of  the
\emph{FWHMs} to  reduce error due to different  seeing \emph{FWHM} but
this  method  is  feasible  only  if  the  frames  involved  in  color
measurement are observed in quick succession. For our observations the
seeing \emph{FWHM} was nearly the same  for the optical $B, V$ and $R$
bands, but significantly better in  $K'$ band as indicated by Table 2.
We have therefore degraded the  $K'$ images to the mean \emph{FWHM} of
the optical  band images  for each galaxy.  The $V-K'$  color profiles
shown in Fig. 2 were obtained using the degraded $K'$ band images.

\begin{deluxetable*}{cccccc}
\tabletypesize{\small} 
\tablewidth{0pt} 
\tablecaption{Color gradients}
\tablehead{ \colhead{Galaxy} & \colhead{r1} & \colhead{r2} &
\colhead{$\Delta (B-V)/ \Delta \log r$} &
\colhead{$\Delta (B-R)/ \Delta \log r$} &
\colhead{$\Delta (V- K')/ \Delta \log r$} \\
\colhead{} & \colhead{(arcsec)} & \colhead{(arcsec)} &
\colhead{} & \colhead{} & \colhead{} }  
\startdata 
UGC 00080 & 3.40 & 28.05 & $-$0.09$\pm$0.01 & $-$0.14$\pm$0.04 & $-$0.22$\pm$0.02 \\
UGC 00491 & 3.40 & 51.00 & $-$0.25$\pm$0.03 & $-$0.26$\pm$0.03 & $-$0.25$\pm$0.00 \\    
UGC 00859 & .... & ..... &    ..... 	      &    .....       &    .....         \\
UGC 00926 & 3.40 & 34.00 & $-$0.21$\pm$0.01 & $-$0.24$\pm$0.01 & $-$0.23$\pm$0.02 \\
UGC 01250 & .... & ....  &    .....           &     .....      & $-$0.78$\pm$0.02 \\ 
UGC 01823 & 3.40 & 34.00 & $-$0.01$\pm$0.00 & $-$0.02$\pm$0.01 & $-$0.20$\pm$0.01 \\
UGC 02039 & 3.40 & 09.35 & $-$0.16$\pm$0.01 & $-$0.23$\pm$0.01 & $-$0.29$\pm$0.00 \\
UGC 03087 & .... & ....  &    .....  	      &    .....       &    .....         \\
UGC 03452 & 3.40 & 09.35 & $-$0.12$\pm$0.00 & $-$0.15$\pm$0.00 & $-$0.40$\pm$0.00 \\
UGC 03536 & 3.40 & 09.35 & $-$0.14$\pm$0.00 & $-$0.19$\pm$0.00 & $-$0.21$\pm$0.00 \\
UGC 03567 & 3.40 & 09.35 & $-$0.09$\pm$0.00 & $-$0.11$\pm$0.00 & $-$0.22$\pm$0.00 \\
UGC 03642 & 3.40 & 13.60 & $-$0.25$\pm$0.00 & $-$0.30$\pm$0.00 & $-$0.20$\pm$0.02 \\
UGC 03683 & 3.40 & 13.60 & $-$0.16$\pm$0.00 & $-$0.16$\pm$0.00 & $-$0.35$\pm$0.05 \\ 
UGC 03699 & 3.40 & 17.00 & $-$0.05$\pm$0.01 & $-$0.12$\pm$0.01 & $-$0.33$\pm$0.01 \\
UGC 03792 & 3.40 & 17.00 & $-$0.34$\pm$0.01 & $-$0.54$\pm$0.01 & $-$0.72$\pm$0.02 \\
UGC 03824 & 3.40 & 21.25 & $-$0.18$\pm$0.01 & $-$0.19$\pm$0.01 & $-$0.06$\pm$0.04 \\
UGC 04347 & 3.40 & 22.95 & $-$0.06$\pm$0.00 & $-$0.09$\pm$0.00 & $-$0.29$\pm$0.01 \\
UGC 04767 & 3.40 & 18.70 & $-$0.12$\pm$0.01 & $-$0.21$\pm$0.01 &  ~~0.12$\pm$0.02 \\
UGC 06389 & 3.40 & 22.95 & $-$0.15$\pm$0.00 & $-$0.23$\pm$0.01 & $-$0.27$\pm$0.01 \\
UGC 06899 & 3.40 & 17.85 & $-$0.19$\pm$0.01 & $-$0.20$\pm$0.01 & $-$0.23$\pm$0.01 \\
UGC 07473 & 3.40 & 17.00 & $-$0.14$\pm$0.00 & $-$0.21$\pm$0.00 & $-$0.42$\pm$0.01 \\
UGC 07880 & 3.40 & 16.15 & $-$0.18$\pm$0.01 & $-$0.23$\pm$0.01 & $-$0.26$\pm$0.01 \\
UGC 08675 & 5.95 & 53.55 & $-$0.09$\pm$0.00 & $-$0.11$\pm$0.01 & $-$0.19$\pm$0.01 \\
UGC 11178 & 3.40 & 17.85 & $-$0.10$\pm$0.01 & $-$0.17$\pm$0.01 & $-$0.29$\pm$0.02 \\
UGC 11356 & 3.40 & 21.25 & $-$0.13$\pm$0.01 & $-$0.19$\pm$0.01 & $-$0.24$\pm$0.01 \\
UGC 11781 & 3.40 & 08.50 & $-$0.12$\pm$0.00 & $-$0.15$\pm$0.00 & $-$0.45$\pm$0.02 \\ 
UGC 11972 & 3.40 & 19.55 & $-$0.08$\pm$0.01 & $-$0.11$\pm$0.01 & $-$0.25$\pm$0.01 \\
UGC 12443 & 3.40 & 14.45 & $-$0.10$\pm$0.00 & $-$0.18$\pm$0.01 & $-$0.16$\pm$0.01 \\
UGC 12655 & 3.40 & 14.45 & $-$0.17$\pm$0.00 & $-$0.24$\pm$0.01 & $-$0.29$\pm$0.01 
\enddata
\end{deluxetable*}

\begin{deluxetable*}{lccc}
\tabletypesize{\small}
\tablewidth{0pt}
\tablecaption{Comparison of color gradients}
\tablehead{
\colhead{} &\colhead{$\Delta (B-V)/ \Delta \log r$} &
\colhead{$\Delta (B-R)/ \Delta \log r$} 
& \colhead{$\Delta (V- K')/ \Delta \log r$} 
}
\startdata
Our sample  &  $-0.13\pm0.06$~($25$)  & $-0.18\pm0.06$~($25$)  & $-0.25\pm0.11$~($25$) \\\\

Ellipticals & $-0.04\pm0.01$~($12$)  & $-0.09\pm0.02$~($30$)  & $-0.16\pm0.19$~($12$)  \\
(Peletier \etal 1990a \& b)&       &                 &                 \\\\ 

Bulges of disk galaxies & ...  &  $-0.19\pm0.16$~($30$) & ...                \\
(Peletier \& Balcells 1997)  &     &                 &                 
\enddata
\tablecomments{The number of galaxies used to obtain the listed mean values 
are indicated in brackets. See $\S$ 4 for details.} 
\end{deluxetable*}

\subsection{Integrated magnitudes}

In this work  we obtain the total magnitude of each  galaxy in all the
bands by  integrating the light  within the isophote  corresponding to
$\mu_{\rm  B}$ =  25~mag\,arcsec$^{-2}$ in  each band.  To  obtain the
$B-V, B-R$  and $V-K'$ colors, however,  we integrate out  only to the
radius at  which the error in $\mu_{\rm  B}$ is 0.1~mag\,arcsec$^{-2}$
(the radius where the  dotted horizontal line intersects the $\mu_{\rm
B}$  profile  in Fig.~2).  The  colors  obtained  in this  manner  are
consistent with those obtained using total magnitude as defined above,
within the estimated errors.  However, the errors on integrated colors
obtained  in this manner  are substantially  lower (typically  0.01 vs
0.04  in $B-V$, 0.01  vs 0.04  in $B-R$,  and 0.05  vs 0.12  in $V-K'$
colors). We  also obtain total  $K'$ magnitudes measured  to $\mu_{\rm
K'}$ =  20~mag\,arcsec$^{-2}$ in order  to make comparison  with 2MASS
photometry.  The measured  total  $B$ magnitudes  and  $B-V, B-R$  and
$V-K'$ colors along  with the estimated errors for  the 29 galaxies in
the sample for which we have  the necessary data are shown in Table 4.
In Fig.  4 (left panel) we  have compared our photometry  with that in
the literature.  We  have plotted (from bottom to  top) the difference
between ~(i-ii)  our $B$ magnitude and  $B-V$ color and  those of RC3,
~(iii) our $B$ magnitude and that  of UGC and ~(iv) our $K'$ magnitude
measured  inside the isophote  $\mu_{\rm K'}$  = 20~mag\,arcsec$^{-2}$
and that of  2MASS, all against our $B$  magnitude.  Agreement between
our $B$ magnitude  with that of RC3 is within  the expected errors for
all  except two galaxies.  Our total  $B-V$ colors  are systematically
around 0.04 magnitude  bluer than those of RC3. On  the other hand, if
we compare the $B-V$ color obtained by integrating only upto $\mu_{\rm
B}$ = 21~mag\,arcsec$^{-2}$ ({\it bulge color}), the agreement between
the colors is  excellent. The agreement between our  $K'$ magnitude is
also  reasonably good.  However it  is  important to  note that  2MASS
magnitudes  correspond to $\mu_{\rm  K'}$ =  20~mag\,arcsec$^{-2}$ and
considerable  flux could  be outside  this radius  \eg our  total $K'$
magnitudes (measured inside the  radius corresponding to $\mu_{\rm B}$
=  25~mag\,arcsec$^{-2}$)  are  systematically brighter  by  $\sim$0.4
magnitudes as  compared to those  of 2MASS.  Hence $K'$  magnitudes in
the 2MASS catalog underestimate the total flux.

To check  the internal consistency  of our photometry we  have plotted
color-color diagrams  as shown  in Fig. 4  (right panel).  In general,
there  is   a  good   correlation  between  these   independent  color
measurements and except  for two galaxies UGC 3087  and UGC 11178, the
estimated   colors  follow  the   sequence  expected   for  early-type
galaxies. Among these, UGC 3087 is clearly outside the 3$\sigma$ error
estimation in all the three colors.  This galaxy is known to harbor an
active galactic  nucleus, which clearly affects the  color profiles in
Fig. 2.   Hence its colors  are expected to depart  significantly from
the general  correlation.  UGC 11178,  though is within  the 2$\sigma$
error bar,  is curious because it  stands out in the  plots, which are
completely  independent  measurements.  Further  photometry  would  be
required to  understand the nature  of this object.  We  exclude these
two galaxies from further analysis.

\subsection{Dust}
We  have  used  optical  and  $K'$  band data  to  study  the  spatial
distribution of  dust. We have  obtained extinction map   in magnitude
$A_{\lambda}$  =  $-2.5~{\rm  log}  (I_{\rm  \lambda,~obs}  /  I_{\rm
\lambda,~model})$, where $I_{\rm \lambda,~obs}$ is the observed image
in same  band, and $I_{\rm \lambda,~model}$ is the  smooth version of
this  image,   obtained  using   the  best  fit   ellipse  parameters.
Extinction map   particularly helping in  identifying non-axisymmetric
structure. Color  maps too  are of help  in identifying  features with
wavelength dependent  intensity.  We have  noticed from the  color and
extinction  maps that  the lenticular  galaxies  UGC\,3178, UGC\,3536,
UGC\,3792, UGC\,4347,  UGC\,7473 and UGC\,7880 show  clear evidence of
the presence  of dust in the form  of lanes or patches.   In all these
cases  there are  differences in  the ellipticity  and  position angle
profiles,  as  well  as  in  the  higher  order  Fourier  coefficients
profiles,  in   the  different   bands.   This  indicates   that  such
differences could  be used as  indicators of the possible  presence of
dust,  even  when the  dust  is  not  otherwise discernible  (see  \eg
Peletier \etal  1990a, Goudfrooij \etal 1994).   The lenticular galaxy
UGC\,3792 has a very prominent dust lane along the major axis (see the
appendix for details).  We will  present elsewhere a detailed study of
the dust in our sample of galaxies.

\plotbig{sb.fig2a}{Surface brightness and color profiles along the major axis
of sample galaxies.For each galaxy, $B$-band surface brightness and the $B-V$, 
$B-R$ and $V-K'$ color profiles (solid lines) are given in separate panels. 
The integrated color of the galaxy in successive elliptical apertures is shown 
by the dotted curve. The galaxy $B$-band intensity corresponding to 10\% 
of the sky brightness (or equivalently an error of 0.1 mag arcsec$^{-2}$) 
is indicated by the dotted line in the top panel. 
The error bars on the surface brightness profiles correspond to 1\% error 
in the estimation of the sky background. Error bars on the color profiles 
are estimated by quadratically adding the uncorrelated part of the errors 
on the two magnitudes that contribute to the color.\label{fig:xxx}}{1}

{\begin{figure*}\epsscale{1.0}
\plotone{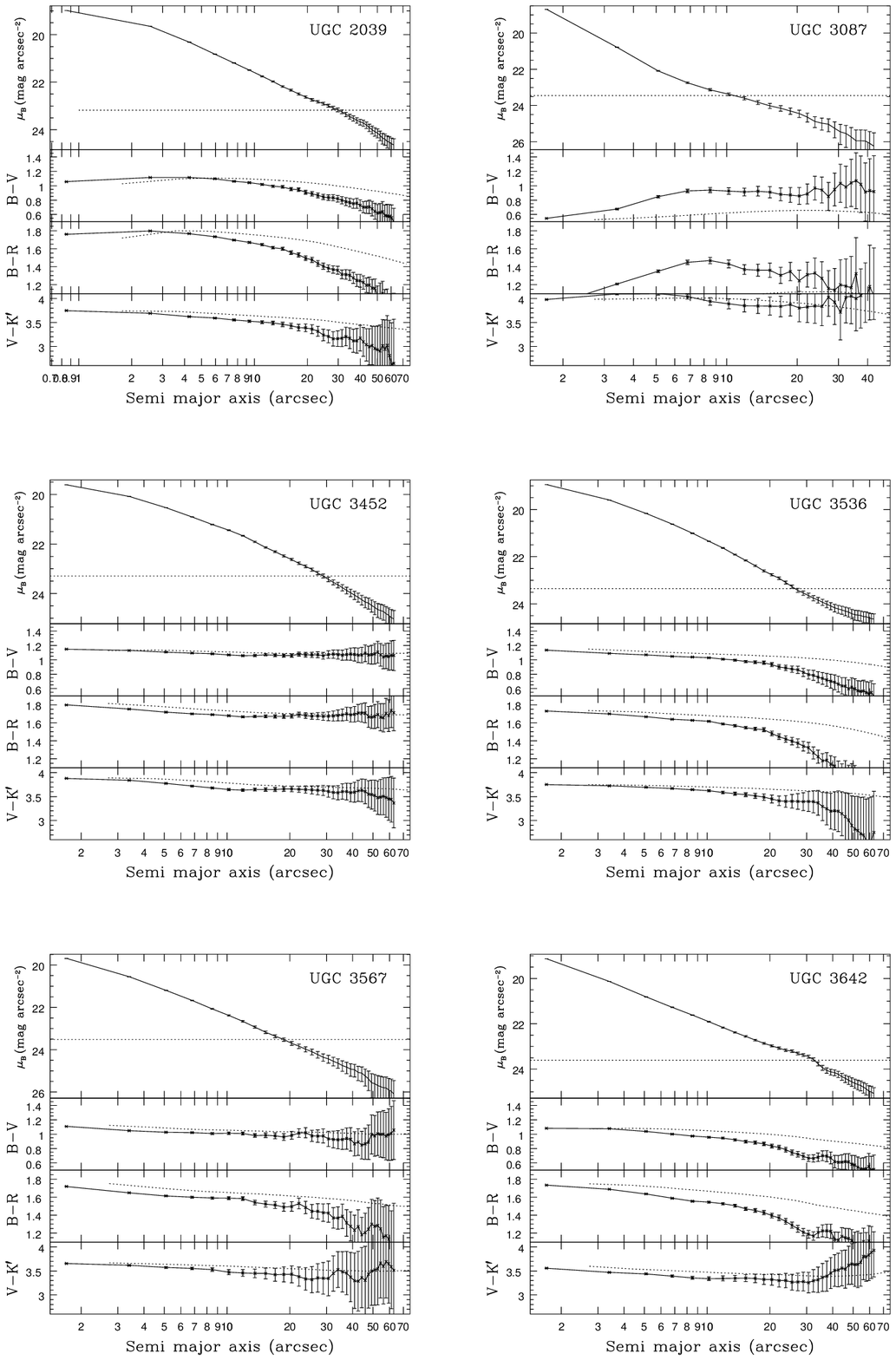}
\begin{flushleft}
FIG.~\ref{fig:xxx} (continued)
\end{flushleft}
\end{figure*}}

{\begin{figure*}\epsscale{1.0}
\plotone{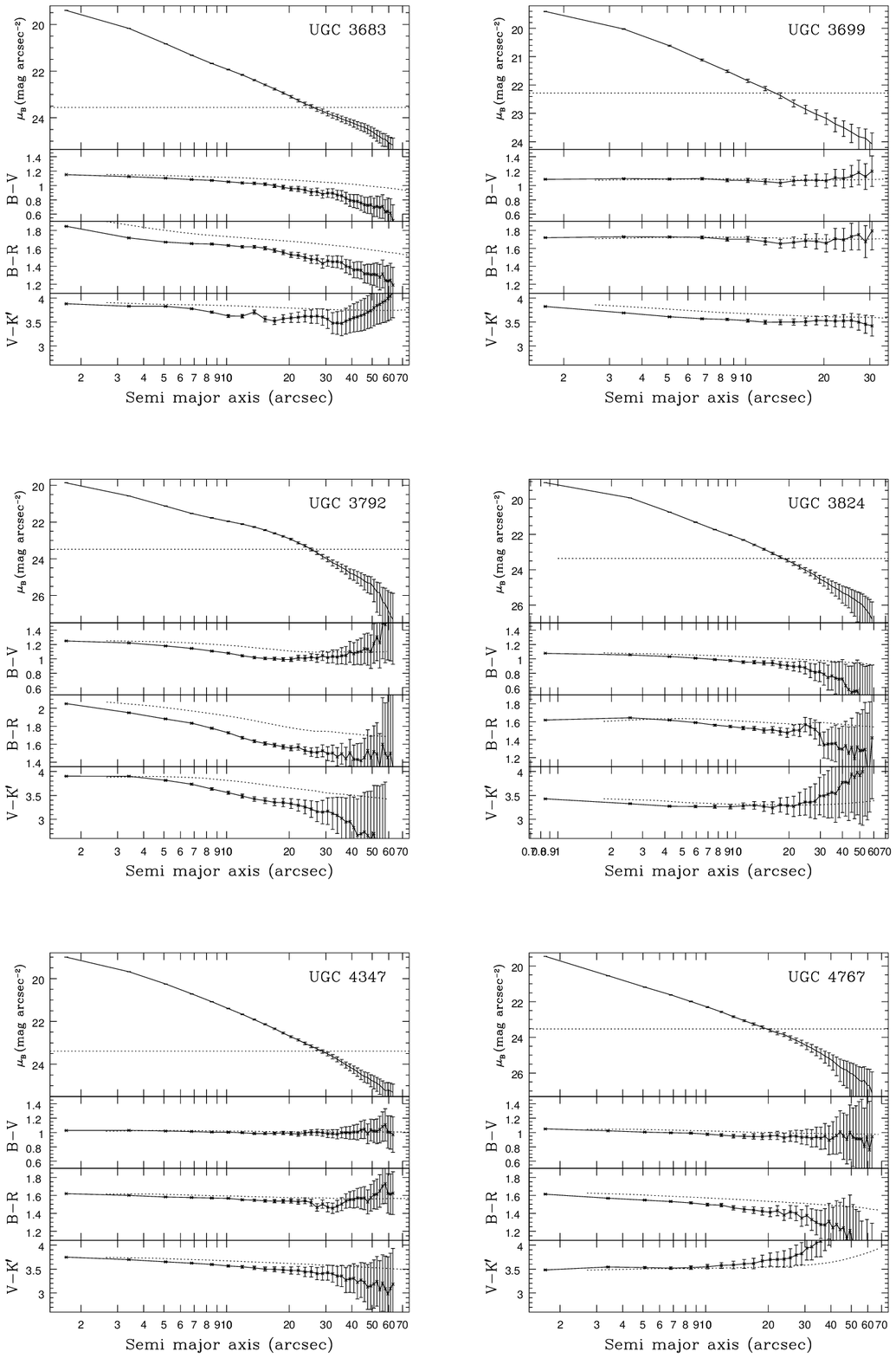}
\begin{flushleft}
FIG.~\ref{fig:xxx} (continued)
\end{flushleft}
\end{figure*}}

{\begin{figure*}\epsscale{1.0}
\plotone{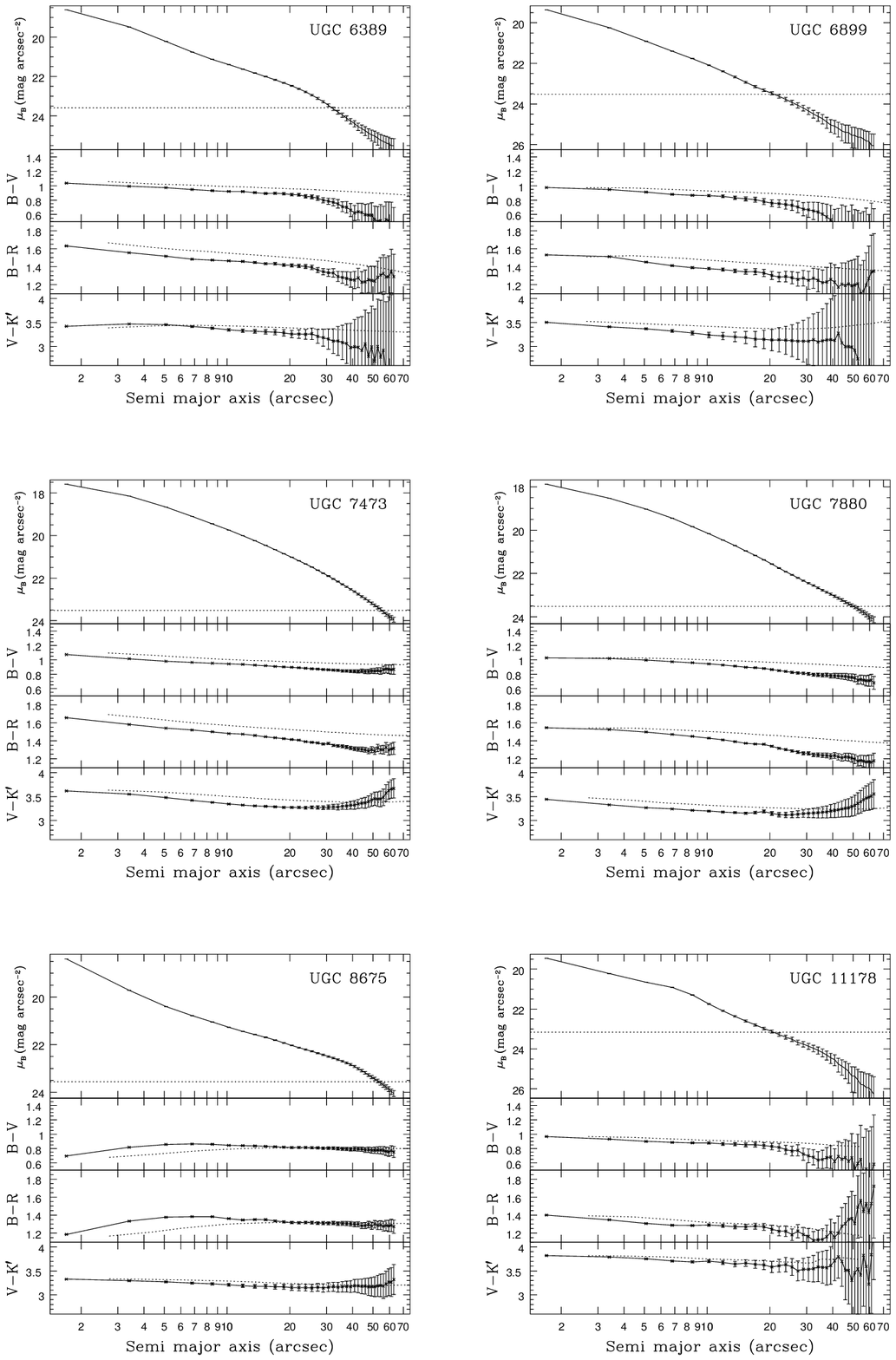}
\begin{flushleft}
FIG.~\ref{fig:xxx} (continued)
\end{flushleft}
\end{figure*}}

{\begin{figure*}\epsscale{1.0}
\plotone{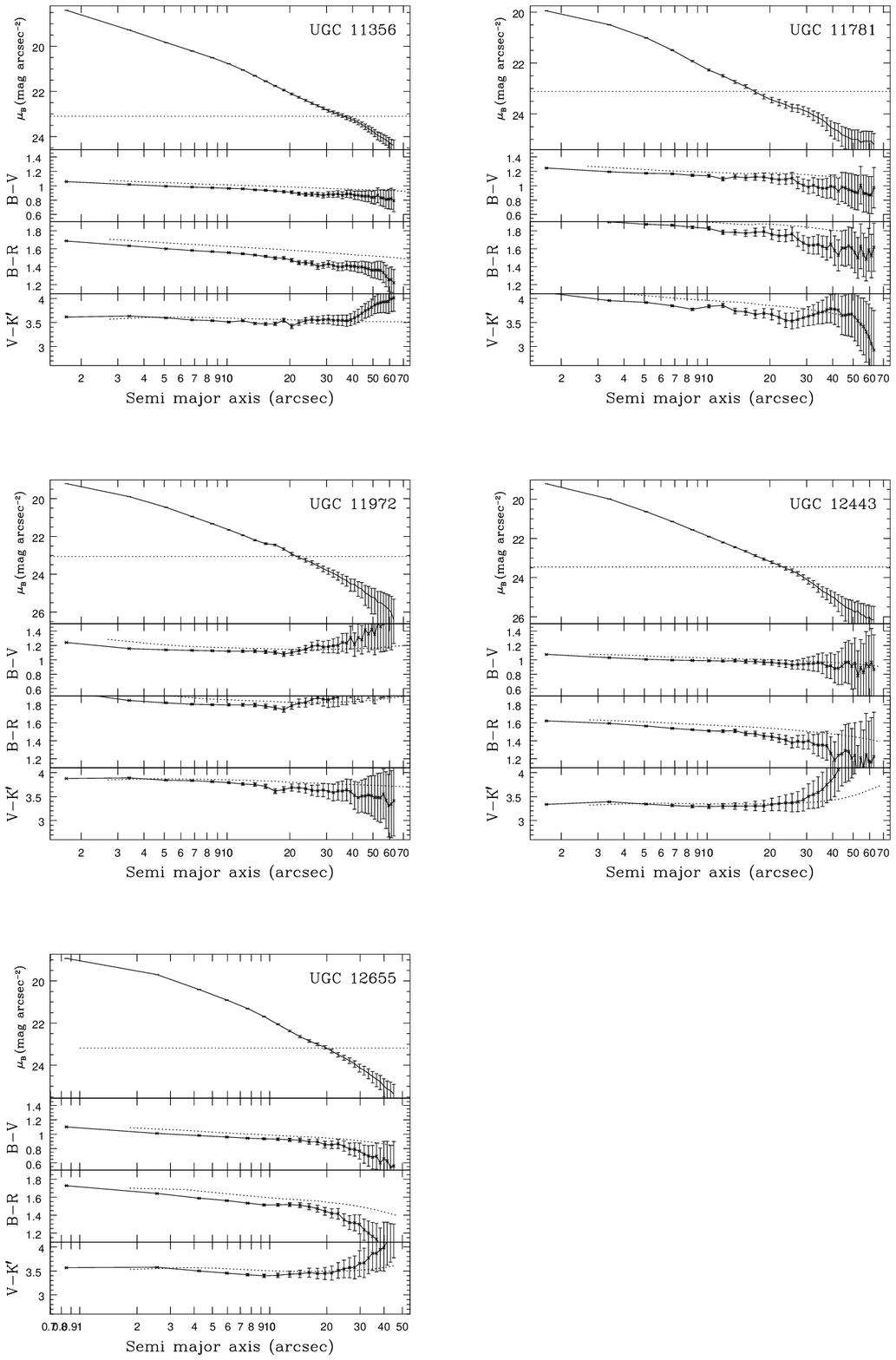}
\begin{flushleft}
FIG.~\ref{fig:xxx} (continued)
\end{flushleft}
\end{figure*}}

\section{Color Gradients}
It is the  usual practice to parameterize the  color gradient over the
galaxy as the slope $\Delta  (B-V)/ \Delta \log r$, which is evaluated
by making a straight line fit  between an inner and an outer radius to
the  color  profiles. The  inner  cutoff limit  (r1)  is  taken to  be
$\sim$1.5 times the seeing FWHM (Peletier \etal 1990a), so that seeing
effects are  lessened. The outer limit  (r2) is taken to  be the point
along the major axis at which the error on the mean surface brightness
of  the fitted  ellipse reaches  $0.1\magarcsecsqi$. It  is  seen from
Fig. 2  that plots  of color against  the logarithm of  the semi-major
axis for  most of our sample  galaxies, have a  reasonably linear form
between the  inner and outer  cutoff. There are some  cases (UGC\,859,
1250 and 3087) for which  the color profiles are non-linear between r1
and r2 because of the presence of a strong ring (UGC\,859, 1250) or an
AGN (UGC\,3087), which makes reliable estimation of the color gradient
is  difficult.  The  $B-V, B-R$  and  $V-K'$ color  gradients for  our
galaxies are given in Table 5.

The colors  of the  galaxies in our  sample become bluer  outwards, in
keeping with the trend observed  in ellipticals. The only exception to
this  is the  positive  gradient in  the  $V-K'$ color  of the  galaxy
UGC\,4767. After excluding the galaxy UGC\,3792 (which has a prominent
dust  lane), the  mean logarithmic  gradients  in the  $B-V, B-R$  and
$V-K'$  colors   are  $-0.13\pm0.06$,  $-0.18\pm0.06$,  $-0.25\pm0.11$
magnitude per  dex in  radius respectively. A  comparison of  our mean
color  gradients with  corresponding  values from  the literature  for
ellipticals and bulges of early-type  spirals is shown in Table 6. The
color gradients  for lenticulars are more negative  than the gradients
for  ellipticals, while  the $B-R$  gradient for  lenticulars  is less
negative  than the  corresponding  gradient for  bulges of  early-type
spirals. Our steeper color  gradients could imply that the metallicity
gradients are stronger in lenticulars and/or that there is more recent
star formation in the outer  parts of lenticulars as compared to those
of   ellipticals.  Bothun   \&  Gregg   (1990)   reported  significant
differences in  the colors  of bulges and  disks in  lenticulars which
they interpret  as being  due to age  difference, and  not metallicity
difference, between the bulge and disk components. The large rms error
in the mean color gradient of our sample could be due to the different
colors  of the  bulge and  disk components  and varying  proportion of
these in different lenticulars.

Correlation  between  different color  gradients  have  been noted  by
various authors (Peletier \etal 1990a \& b and Idiart \etal 2002 ) and
have been used  to derive information about the  physical processes in
galaxies. Such correlations are displayed in Fig. 5 for our sample.  A
clear  correlation is  seen between  $B-V$ and  $B-R$  gradients, with
correlation coefficient 0.90 at a significance level better than 99.99
percent. Correlation between $B-V$ and $B-R$ gradient has been noticed
for a  sample of  ellipticals by  Peletier \etal (1990  a \&  b). From
their  data  we  obtain  a   correlation  coefficient  of  0.97  at  a
significance  level better  than 99.99  percent.  The  plot  of $V-K'$
against $B-V$  gradients shows  large scatter, and  we do not  see any
correlation of  the type  reported by Peletier  \etal (1990b),  but it
should  be  noted  that they  used  only  13  galaxies to  derive  the
correlation. It is necessary to investigate how the separate bulge and
disk color  gradients affect the  total color gradients and  hence the
correlation between them.  We will address in the  future these issues
using  a bulge-disk  decomposition technique  which will  allow  us to
study the color distribution in these components separately.

\plotbig{sb.fig3}{Isophotal profiles for the galaxy UGC\,7880.  See the
text for details.}{1}

\section{Conclusions}
In this paper we have presented detailed multicolor surface photometry
performed with a CCD  in the $B, V, R$ bands and  a NICMOS detector in
the $K'$  band, for a  sample of 34  lenticular galaxies from  the UGC
catalogue.  The  galaxies were  chosen in an  unbiased fashion  from a
subset of UGC lenticulars as explained in Section 2.  We have obtained
total  integrated magnitudes  and colors  for all  the  galaxies using
elliptical annuli from surface photometry,  and find that these are in
good agreement  with values from  the RC3 catalogue.   Using isophotal
analysis we  have obtained radial profiles of  the surface brightness,
ellipticity, position angle, and  higher order Fourier coefficients in
all the  bands.  The  profiles in the  different bands  are consistent
with each other, and any differences can be attributed to the presence
of  dust  and  other   features  which  produce  wavelength  dependent
effects. We have used the  surface brightness profiles to obtain color
profiles, and logarithmic color gradients, and find that the gradients
are negative, indicating that  the colors of lenticulars become redder
towards the center, as is  the case with elliptical galaxies.  We have
shown that  there is  good correlation between  $B-V$ and  $B-R$ color
gradients for lenticulars.  Numerical  profiles of all parameters that
we  have obtained  from the  isophotal analysis  along with  the color
images            are           available            at           {\it
http://www.iucaa.ernet.in/$\sim$sudhan/s0.html}.

Our intention in obtaining the  multiband data on lenticulars has been
to study  in detail these  galaxies as a  class, and to  compare their
properties  with  those of  ellipticals  and  early  type spirals.  An
important  aspect of  this  study  will be  the  decomposition of  the
lenticulars into bulge and disk components and the comparison of these
separately with the bulges and  disks in other types of galaxies where
these components  occur with varying  degrees of prominence.   We will
use  the results  of the  decomposition in  a multiband  study  of the
fundamental and photometric planes for lenticulars.  We will also make
a  detailed  study  of   the  distribution  of  dust  in  lenticulars,
particularly  in  those  galaxies  from  our sample  where  there  are
prominent dust features and where we have multiband data at optical as
well as near-infrared wavelengths.

\acknowledgments 
The staff at OAGH and SPM are greatfully acknowledged
for their  help during  the observations. SB  and SKP thank  IUCAA for
hospitality and  the use of  facilities without which this  work could
not have been  done and for providing funds  for observations. AKK and
SB thank INAOE for hospitality  provided during there visits.  We also
thank the anonymous referee  for several useful comments, which helped
to improve the original manuscript.  This research has made use of the
NASA/IPAC Extragalactic  Database (NED), which is operated  by the Jet
Propulsion  Laboratory,  California  Institute  of  Technology,  under
contract with the National Aeronautics and Space Administration.

\appendix 
In  this appendix, we discuss structural  properties of each
galaxy, whenever isophotes  in one or more bands  depart from a smooth
ellipse.   The discussions  are  focused on  any  disagreement in  the
morphological classification in UGC and RC3 catalogs, presence of dust
lane,  change  of  ellipse-fit  parameters (b4,  PA  or  ellipticity),
existence of any ring, and any evidence of nuclear activity. Published
information on any  of these issues, if available,  is noted. In total
12 galaxies (UGC\,926, 1823, 3452, 3567, 3683, 3699, 3824, 6013, 6389,
6899,  11972 and  12443) are  smooth lenticular  galaxies  without any
identifiable  feature in  the  direct images,  the surface  brightness
profiles, color or extinction maps. Comments for the remaining
galaxies follow.

\noindent{\bf UGC\,80 :} UGC\,80 is a barred lenticular galaxy, having
a  very   faint  interacting  companion.  The  faint   bar  is  almost
perpendicular to the galaxy major  axis, and is clearly evident in the
isophotal profile. The  $b_4$ profile is indicative of  a disky shape
over  the  entire  major-axis  range explored.   The  ellipticity  and
position  angle profiles  show  abrupt change  at $r\simeq  5\arcsec$,
where there is also a kink  in the brightness profile. The $B-R$ color
map of the galaxy shows a red feature near the center.

\noindent{\bf UGC\,491 :} UGC\,491 is  the brightest galaxy in a small
group with NGC\,0258 and NGC\,0260.  This galaxy is classified as S0 in
the UGC catalogue and SA(r) in the RC3 catalogue. In the optical images it
indeed shows  a ring, which  is also evident  in the $B-R$  and $B-K'$
color  maps. Ellipticity  and  position  angle  profiles show  a  change  
at $r\sim21.5\arcsec$, which may be due to the presence of the ring.

\noindent{\bf UGC\,859 :}  UGC\,859 shows an internal ring,  but it is
classified  as S0 in  the UGC  catalogue. The  $B-R$ and  $B-K'$ color
maps of the  galaxy show the ring as  an inhomogeneous structure.  A
low surface brightness disk with some luminous spiral structure exists
outside the  ring.  UGC  859 has  been detected by  IRAS and  has flux
densities  of $1160\pm33\mjy$  at $60\micron$  and  $870\pm319\mjy$ at
$100\micron$ (Knapp \etal  1989).   The $b_4$ coefficient  is disky
towards the outer region.

\noindent{\bf  UGC\,1250 :}  While this  galaxy is  classified  as S0,
the $B-R$ color map of the galaxy  show a very clear, but distorted ring,
just around the bulge. $\halpha + [NII]$ observations by  Pogge \& Eskridge  
(1993) show copious HII  regions across  the  disk.  The  $b_4$  profile 
is  disky at  all major-axis  lengths, with  the coefficient  being larger  
in  the inner region of the galaxy.

\noindent{\bf UGC\,2039 :} UGC\,2039  is paired with UGC 2049. There is
a rise  in the ellipticity from  0.1 to 0.5 towards  the outer region.
The $b_4$  coefficient is disky  in the outer region,  which suggests
the presence of  a significant disk.  The galaxy  has been detected by
IRAS at 60 and $100\micron$.
\plotbig{sb.fig4}{Comparison  of our  photometry with  those in  the literature.
Left panel (from  bottom to top): Difference between  our $B$ and that
of RC3, our $B-V$  and that of RC3, our {\it bulge}  $B-V$ and that of
RC3, our $B$ and  that of UGC and our $K'$ and  that of 2MASS, plotted
against  our B  magnitude.  Error  bars on  our data  (left),  and the
literature data  (right) are plotted  on the difference=0  line. Right
panel:  Color vs color  plots from  our data.  Typical error  bars are
indicated on the bottom right part of the figure.}{1}
\noindent{\bf UGC\,3087  :} Previous studies of UGC\,3087 have mainly 
concentrated on its nuclear activity. It is  a strong  radio source 
(3C120) and the  optical  spectrum suggests a Seyfert 1 nucleus  
(Tadhunter \etal 1993). It  has  a faint  optical jet  in the  same
apparent  direction as  the  radio  jet. The  optical  jet is  clearly
visible in  our images  and distorts isophotal  profiles in  the inner
region.

\noindent{\bf UGC\,3178 :} A dust  feature is clearly visible in the $B$
band  extinction map of  this galaxy,  and is  also detectable  from the
isophotal analysis of $B$ and $K'$ band images. The $b_4$ coefficient,
position angle  and ellipticity are  different in $B$ and  $K'$, which
may be  due to dust  absorption in $B$.   Our $V$ and $R$  band images
suffer from poor S/N.

\noindent{\bf UGC\,3536  :} This galaxy  has a significant  dusty disk
and the $b_4$ coefficient is disky.   The disk is also evident in the
$B-V$ and  $B-R$ color maps. There  is a constant  rise in ellipticity
for  semi-major axis length  $<19.5\arcsec$, beyond  which there  is a
decrease, which could be because of the disk.

\noindent{\bf UGC\,3642 :} The images for this galaxy show  some kind 
of spiral structure, but this is not evident  in the extinction or color 
maps. The $b_4$ coefficient is positive in the outer regions.

\noindent{\bf UGC\,3792 :} This galaxy  is classified as S0 in the UGC
but  as SA0/a  in  the RC3  catalogue,  and has  no previous  reported
photometric  study. The  galaxy has  a prominent  dust lane  along the
major axis which greatly affects  the various profiles. This dust lane
is most prominent  in the $B$ band. The  $b_4$ coefficient is positive
in the  outer region.  There is  a large color gradient  in $B-V, B-R$
and $V-K'$,  presumably because of the  dust. A detailed  study of this
galaxy in the $B, V, R, J, H, K'$ bands  will be presented in a forthcoming
paper.
{\begin{figure*}
\begin{center}
\includegraphics[scale = 1, width = 0.8\textwidth,trim = 0 250 30 0, clip]{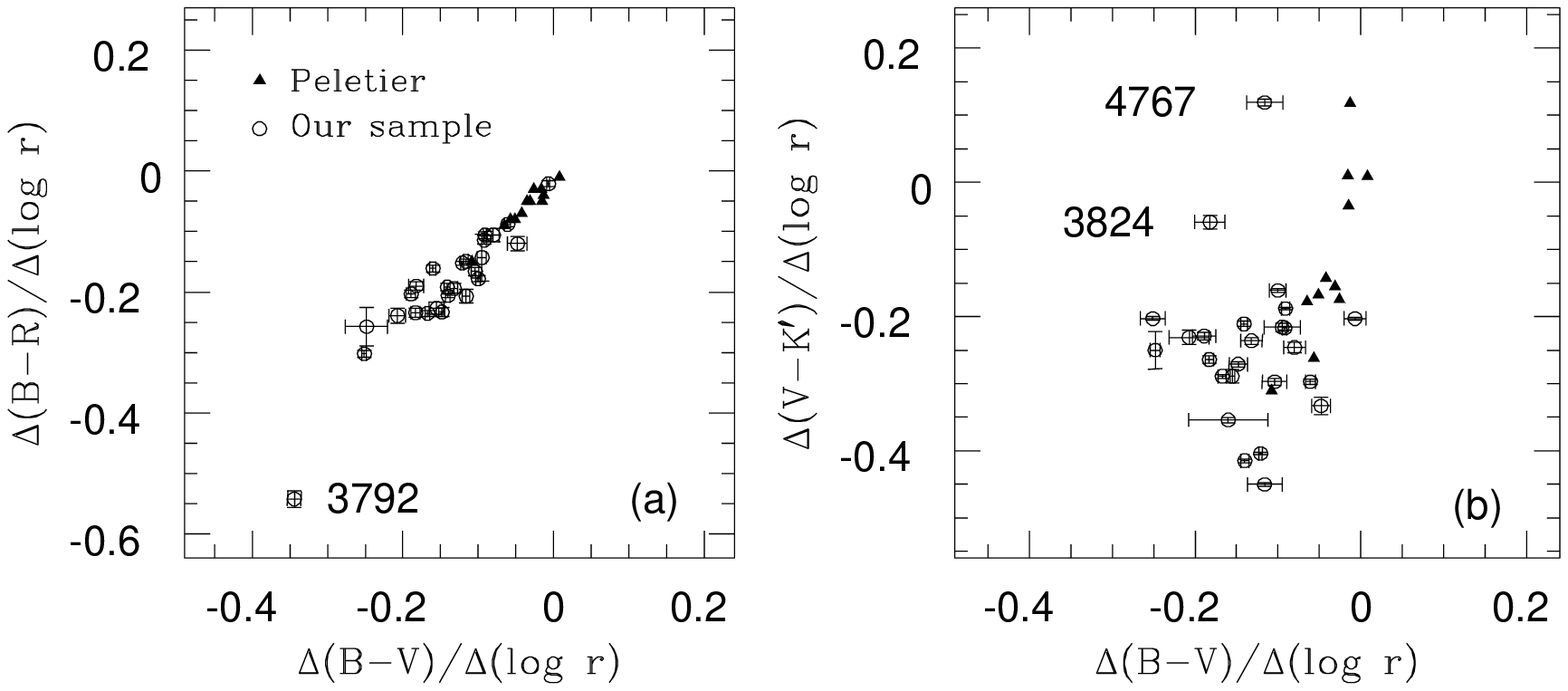}
\caption{Correlation between  different color gradients.  The gradients
for  our sample  and the  sample of  elliptical galaxies  of Peletier
\etal (1990a \&  b) are indicated with different  symbols as in panel
(a).}

\end{center}
\end{figure*}}

\noindent{\bf UGC\,4347  :} Our $B-V$ and $B-K'$  color and extinction
maps of  this galaxy reveal a  large dust patch near  the center.  The
$b_4$  profile is  disky throughout  the observed  region.   Forbes \&
Thomson  (1992) have  detected  possible shells  in  this galaxy.  The
extinction images  in all bands  shows faint structures, which  may be
responsible for the non-zero values the of $b_4$ coefficient.

\noindent{\bf UGC\,4767 :} The  position angle and ellipticity profile
for  this galaxy  are  different in  the  $B, R$  and  $K'$ bands  for
semi-major axis lengths $<5\arcsec$, which  could be because of a dust
patch  near the  center of  the galaxy.  The various  color  maps also
indicate the presence of the dust patch.

\noindent{\bf UGC\,4901  :} Our $B$ band  image of this  galaxy has poor
S/N.  The  color maps in  $V-R$ and $V-K'$  show no features  in the
galaxy. The  $b_4$ coefficient is  positive towards the  outer region,
suggesting the presence of a faint disk.

\noindent{\bf UGC\,7473 :} This well studied edge-on lenticular galaxy
forms a  pair with NGC 4340.   The surface brightness  profile of this
galaxy clearly  indicates the presence  of bulge and  disk components.
The Mg$_2$  line strength profiles along  the major and  minor axis of
the galaxy differ dramatically  and convincingly indicate the presence
bulge  and  disk  components  (Fisher,  Franx  \&  Illingworth  1996).
Michard \& Marchal  (1993) described UGC\,7473 as having  a disk fully
embedded  in a spheroidal  halo.  A  disk of  rapidly rotating  gas is
present  within  the inner  $3\arcsec$;  this  is  decoupled from  the
stellar component (Fisher 1997).  A  concentration of dust in the disk
has been proposed by Michard  \& Poulain (2000).  Our $B-R$ and $B-K'$
color  and extinction  maps also  reveal a  clear inclined  disk.  The
ellipticity rises  from 0.2 to 0.61  over the observed  region and the
$b_4$  coefficient becomes  significantly positive  beyond $6\arcsec$,
which reflects the presence of a strong disk.

\noindent{\bf UGC\,7880  :} This  galaxy forms a  pair with  NGC 4635,
with a  separation of at  $1.5\arcmin$.  The ellipticity  changes from
0.1 to  0.5 upto  $10\arcsec$, and decreases  beyond that.   The $b_4$
coefficient is positive upto the point at which the ellipticity begins
to decrease.   This may be  due to the  presence of an inner  disk, the
presence  of which is  also indicated  by the  $B-R$ and  $B-K'$ color
and extinction maps.

\noindent{\bf UGC\,7933 :} This galaxy is classified as S0 in the UGC,
but as E1-2 in the RC3 catalogue. It forms a non-interacting pair with
NGC\,4670 at  $5.6\arcmin$ separation.  The  $B$ band  image has  poor
S/N. The $b_4$ coefficient is  positive towards the outer region.  The
color and extinction maps do not indicate any structure.

\noindent{\bf UGC\,8675 :} This is  a little studied S0 galaxy hosting
a Seyfert 1.5 nucleus, and has  been classified as a SA0 galaxy in the
RC3 catalogue. The isophotes of  the galaxy are nearly circular and at
$5\arcsec$ the  ellipticity becomes 0.15 which is  consistent with the
values obtained by De Robertis,  Hayhoe \& Yee (1998) and Pierre \etal
(2000).  A  dust absorption  pattern is seen  near the nucleus  in the
$B-R$ and $B-K'$ color and  extinction maps.  Pierre \etal (2000) have
found from HST  observations a U-shaped dust lane  circling around the
nucleus.

\noindent{\bf  UGC\,11178  :} This  galaxy has  no  previous reported
photometric study. The  extinction map  show  a  ring  like  structure
in  all the  bands. The  ellipticity  abruptly changes  between 5  and
$7\arcsec$, which  could be  because of the  ring like  structure. The
$b_4$ coefficient is significantly positive upto $7\arcsec$. The color
maps are featureless.

\noindent{\bf UGC\,11356  :} This is  a face-on lenticular  galaxy.  A
faint structure  is apparent in  extinction images in the  optical bands
near the  center of  the galaxy,  but no features  are obvious  in the
color maps.

\noindent{\bf UGC\,11781 :} This galaxy is classified as S0 in the UGC
but as SAB in the RC3 catalogue. The presence of a bar is clear in the
direct  as well as  extinction images.   There is  a dip  in ellipticity
profile at $10\arcsec$, where there  is a change in the position angle
profile as well. The $b_4$ coefficient is positive in this region.

\noindent{\bf UGC\,12655  :} The $B-R$  and $B-K'$ color  and extinction
maps  of this  galaxy show  a patchy  region near  the center  of this
galaxy. The $b_4$ coefficient is positive upto $8\arcsec$.

\end{document}